\documentclass{WileyMSP-template}

\begin{document}

\pagestyle{fancy}
\rhead{\includegraphics[width=2.5cm]{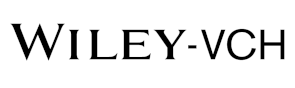}}

\title{Encapsulation of the Graphene Nanoribbon Precursor \\ 1,2,4-trichlorobenzene in Boron Nitride Nanotubes at \\ Room Temperature}

\maketitle


\author{Ana Cadena*}
\author{\'Aron Pekker}
\author{Bea Botka}
\author{Erzs\'ebet Dodony}
\author{Zsolt Fogarassy}
\author{B\'ela P\'ecz}
\author{Katalin Kamar\'as*}


\dedication{ }

\begin{affiliations}
Ana Cadena$^*$\\
Institute for Solid State Physics and Optics, \\
Wigner Research Centre for Physics, \\
P.O. Box 49, 1525 Budapest, Hungary \\
Department of Chemical and Environmental Process Engineering, \\
Budapest University of Technology and Economics, \\
M\H uegyetem rkp. 3, 1111 Budapest, Hungary \\
Dr. \'Aron Pekker, Dr. Bea Botka, Prof. Katalin Kamar\'as$^*$ \\
Institute for Solid State Physics and Optics, \\
Wigner Research Centre for Physics, \\ 
P.O. Box 49, 1525 Budapest, Hungary \\
Email Address: anacristina.cadena@wigner.hu,kamaras.katalin@wigner.hu

Erzs\'ebet Dodony, Dr. Zsolt Fogarassy, Prof. B\'ela P\'ecz \\
Institute of Technical Physics and Materials Science, \\
Centre for Energy Research, \\
P.O. Box 49, 1525 Budapest, Hungary \\

\end{affiliations}



\begin{abstract}

Graphene nanoribbons are prepared inside boron nitride nanotubes by liquid phase encapsulation and subsequent annealing of 1,2,4-trichlorobenzene. The product is imaged with high resolution transmission electron microscopy, and characterized by optical absorption and Raman spectroscopy. Carbon-containing material is detected inside the boron nitride nanotubes with energy-dispersive x-ray spectroscopy (EDS) and scanning transmission electron microscopy (STEM). The observed structures twist under the electron beam and the characteristic features of nanoribbons appear in the Raman spectra.

\end{abstract}


\section{Introduction}

 Hollow nanostructures, being able to encapsulate molecules in their cavities, show considerable application potential from sensing to protective storage. They also offer a route to modify chemical reactions by restricting the reaction space, thereby determining the possible products. Such templated synthesis can also result in the growth of one-dimensional nanostructures, such as graphene nanoribbons.
 
 The cavities of hollow nanomaterials have been used to accommodate molecules,\cite{Smith98, Yudasaka03} which can be further processed to obtain one-dimensional products with controlled size.\cite{Talyzin11, Lim13} In this regard narrow graphene nanoribbons (GNRs), less than a nanometer wide, have been extremely popular targets.\cite{Kitao20, kuzmany21} Single-walled carbon nanotubes (SWCNTs) are the natural choice for this purpose. They are available in a broad variety of diameters, which is convenient to form a range of products, as the size of the assembled ribbons depends on both the precursor and the carbon nanotube inner diameter. SWCNTs are thermally stable in vacuum up  to 1400 $^{\circ}$C, therefore high temperature processing is not a limiting factor to obtain the desired product. A drawback of this technique is that GNRs encapsulated in SWCNTs (GNR@SWCNTs) are generally not directly suitable for applications, such as field effect transistors, but have to be transferred to insulating substrates\cite{Borin19} or directly grown on such surfaces.\cite{Mutlu21} The extraction of nanoribbons has not yet been elaborated. To overcome this problem, metal-organic frameworks (MOFs) have been explored for GNR production, where the extraction of armchair graphene nanoribbons (AGNRs) was accomplished.\cite{Kitao20} This method is limited in temperature up to 500 $^{\circ}$C by the thermal stability of the MOF host.\cite{guillerm12}
 
 Boron nitride nanotubes (BNNTs)\cite{Chopra95} are insulators with a much larger bandgap than even semiconducting carbon nanotubes. Their thermal stability is comparable to SWCNTs,\cite{chen2004} thus they support thermal transformations of the encapsulated species. Their $\pi$-electron density is, however, lower than that of  carbon nanotubes, leading to weaker $\pi$-$\pi$ interactions with aromatic molecules. As a consequence, it is easier to extract the reaction products than in the case of carbon nanotubes.\cite{walker17} Additionally, BNNTs are transparent in the visible range, hence reactions inside the nanotubes can be followed by optical methods. 
 
\begin{figure}
\centering
\includegraphics[width= 0.7 \linewidth]{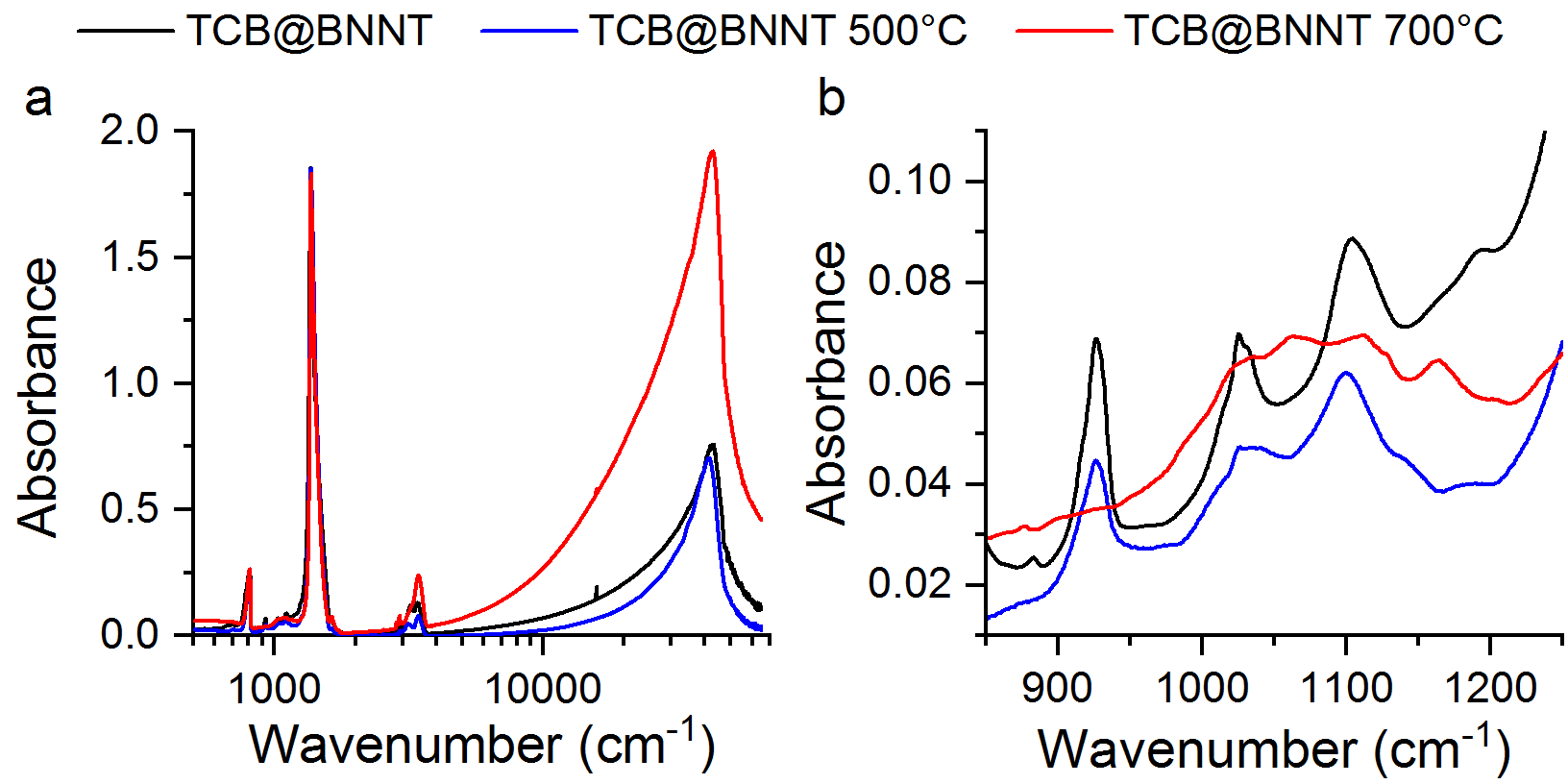}
\caption{a) Wide-range optical absorption spectra of TCB filled BNNT (black) compared to the 500$^{\circ}$C (blue) and 700 $^{\circ}$C (red) annealed samples. Note the logarithmic frequency scale. b) The infrared region, containing TCB molecular vibrations that gradually disappear upon annealing.}
\label{Figure_1}
\end{figure}
 
 The first molecular species to be encapsulated in BNNTs was C$_{60}$ and the method was sublimation.\cite{Mickelson03} An outlook on boron nitride nanotubes was presented in Ref.\cite{cadena21}. After encapsulation, C$_{60}$ molecules can be removed from the interior of the nanotubes by sonication in toluene.\cite{walker17} This also means that care has to be taken when removing the excess of guest species after filling: mild conditions have to be used to prevent extraction from the cavity. Recently, BNNTs have been used to encapsulate dye molecules which are often toxic and protect them from photodegradation.\cite{Allard20}
 
 Here we show that a common solvent, 1,2,4-trichlorobenzene (TCB), can be used as a precursor for growth of GNRs inside BNNTs. As we discussed in our recent work \cite{Cadena22}, where GNRs were prepared inside carbon nanotubes (GNR@SWCNT) starting from  TCBliquid TCB, this molecule has several advantages over larger polycyclic aromatic ones. As it is a liquid at ambient conditions, the filling can be performed by simply immersing the host tubes into the solvent, resulting in a high filling ratio. The liquid precursor also alleviates the problem of purifying the exterior of the tubes, as the excess can be removed by letting it evaporate. Because the filling is performed at room temperature, the encapsulation step can be kept separate from the polymerization and subsequent nanoribbon growth.\cite{Barzegar16}

\section{Results and discussion}

The encapsulation step was performed at room temperature after opening the BNNTs by immersion in liquid TCB. For annealing temperatures we chose 500 $^{\circ}$C (TCB@BNNT 500$^{\circ}$C) and 700$^{\circ}$C (TCB@BNNT 700 $^{\circ}$C), respectively, since shorter and longer ribbons were observed setting these parameters in SWCNT analogues.\cite{Cadena22} 

\begin{figure}[ht]
\centering
\includegraphics[width= 1 \linewidth]{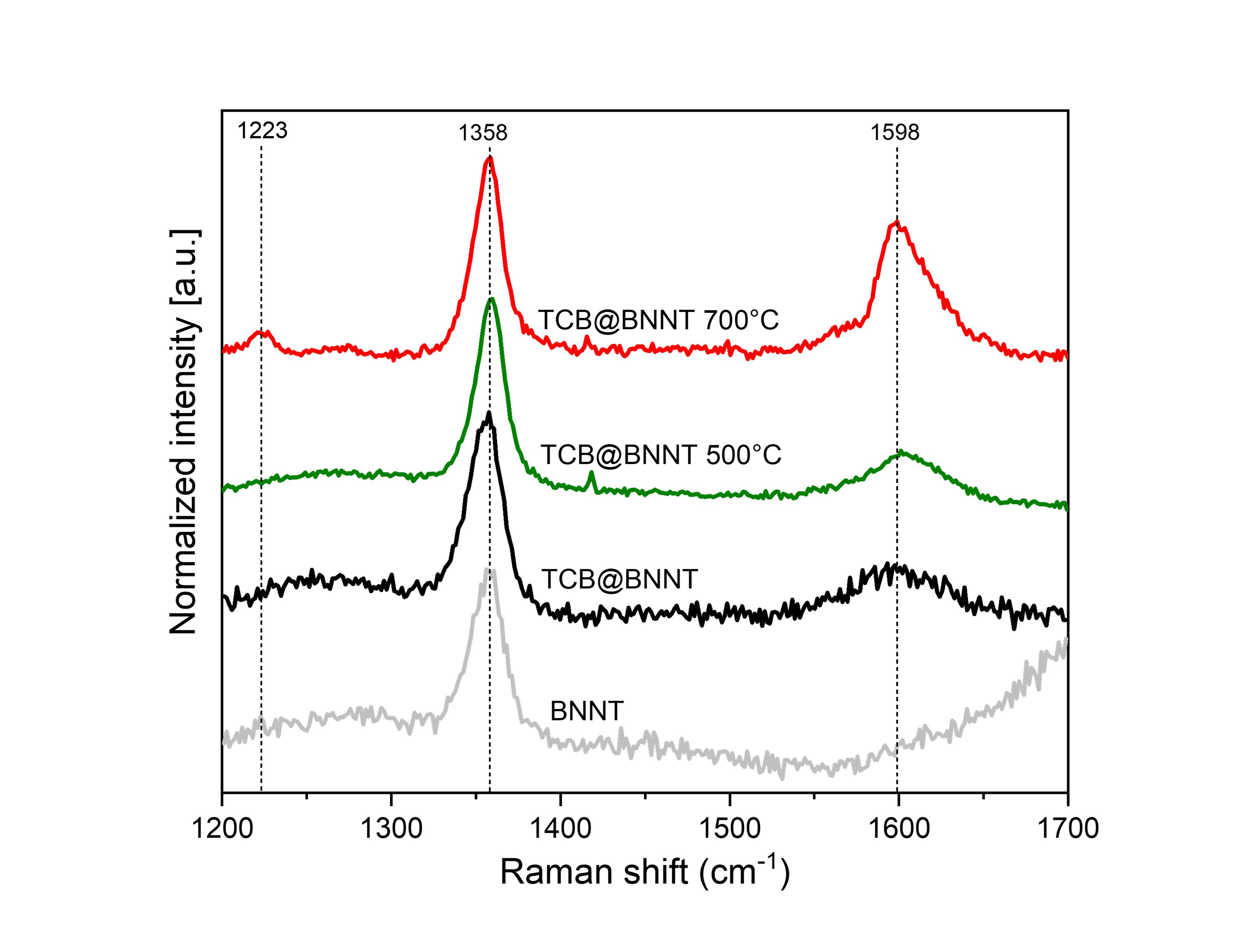}
\caption{Raman spectra of the GNR@BNNT samples compared to  TCB@BNNT and BNNT (as-purchased). A peak at 1223 cm$^{-1}$ appears in the material formed at 700 $^{\circ}$C, comparable to the one observed in other ribbon-like structures. Because of the high luminescence of BNNTs, UV excitation (355 nm) was used. All spectra were background corrected.}
\label{Figure_2}
\end{figure}

The as-received BNNT samples appear as white powder as expected from a wide bandgap insulator ($E_{BG}=$ 5.7 eV\cite{BNNT-LLC}). After filling and filtering, the sample (TCB@BNNT) assumes a grey tone, which disappears after letting the adsorbed molecules on the outside walls evaporate. Subsequent annealing results in a silver/black colored product. Wide-range absorption spectra (\textbf{Figure \ref{Figure_1}a}) are dominated by scattering in the NIR-UV range. This strong background covers the absorption features responsible for the dark color. In the infrared region, mainly boron nitride modes are present; however, in TCB@BNNT small vibrational bands appear around 1000 cm$^{-1}$ (1027, 1105 and 1192 cm$^{-1}$) (\textbf{Figure \ref{Figure_1}b}) that are absent in the 700 $^{\circ}$C annealed sample (although still present after annealing at 500 $^{\circ}$C). These values do not match exactly the TCB modes but are very close to those reported by Pei et al.\cite{Pei13} in TCB adsorbed on graphene. They can be assigned to the encapsulated TCB molecules, observable through the transparent nanotube walls, and their disappearance indicates the progress of the chemical reaction leading to nanoribbons. 

\begin{figure}
\centering
\includegraphics[width= 0.3\linewidth]{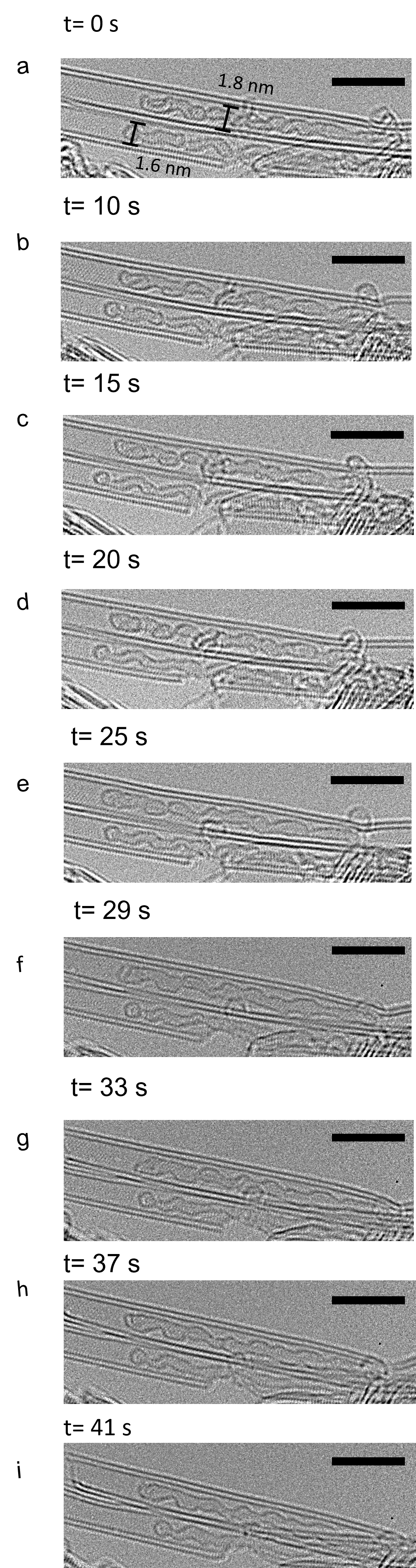}
\caption{Low dosage HRTEM time series of GNR@BNNT captured at 480kx magnification with 336 msec exposure time for each image at parallel illumination with 80kV acceleration voltage. The length of the scale bar corresponds to 5 nm. }
\label{Figure_3}
\end{figure}

The BNNT, filled and annealed samples were all characterized by Raman spectroscopy with excitation by multiple laser sources (see \textbf{Figure S1,} Supporting Information). In most cases, the Raman spectrum was buried under the photoluminescence background originating from color centers present in defective or strained boron nitride structures.\cite{Ciampalini22} The 355 nm laser excitation generated the weakest photoluminescence (\textbf{Figure \ref{Figure_2}}). The BNNT sample shows the well known $E_{2g}$ mode of BNNTs at 1358 cm$^{-1}$. Apart from this peak the spectrum contains no other features. In the spectrum of the filled sample (TCB@BNNT) the characteristic G-mode of sp$^2$ hybridized carbon appears at 1598 cm$^{-1}$. The width of this band decreases on annealing, suggesting the formation of more extended structures at elevated temperatures. In the spectrum of the 700 $^{\circ}$C sample a new mode at 1223 cm$^{-1}$ emerges. A mode in this frequency region has been assigned to a C-H in-plane bending mode in ribbon-like structures \cite{kuzmany21,Borin19,Mutlu21,Gillen09}. We regard the appearance of this mode as proof for the presence of nanoribbons, although neither the  low-frequency radial breathing-like mode (RBLM, masked by the photoluminescence background) nor the D mode (covered by the BNNT band) can be identified in the spectra. The 1223 cm$^{-1}$ band is not observed in the sample prepared at 500 $^{\circ}$C, contrary to what was found in carbon nanotubes.\cite{Cadena22} We note that polymerization of polycyclic hydrocarbons happens at lower temperature at carbon surfaces\cite{Talyzin11} than in the neat materials.\cite{Talyzin11c} The higher temperature where this spectral feature appears indicates that boron nitride constitutes a neutral surface with no catalytic effect. The prevailing intensity of the TCB vibrational bands in the infrared absorption spectra (Figure \ref{Figure_1}) also points to the reaction being incomplete at this temperature. 

Transmission electron microscopy images of the TCB@BNNT 700 $^{\circ}$C sample revealed multi-walled (2-6) BNNTs, in most cases partially filled with material containing carbon, as energy-dispersive x-ray spectroscopy (EDS) and scanning transmission electron microscopy (STEM) maps using EDS confirm. Carbon was identified by STEM-EDS measurements  in the areas where encapsulated material was found (\textbf{Figure S2}, Supporting Information). Selected carbon structures could be identified that moved under the 80 keV electron beam, clearly visible on the time-lapse HRTEM images (\textbf{Figure \ref{Figure_3}}), exhibiting the  characteristic twist of GNRs. 

\begin{figure}[ht]
\centering
\includegraphics[width= 0.9 \linewidth]{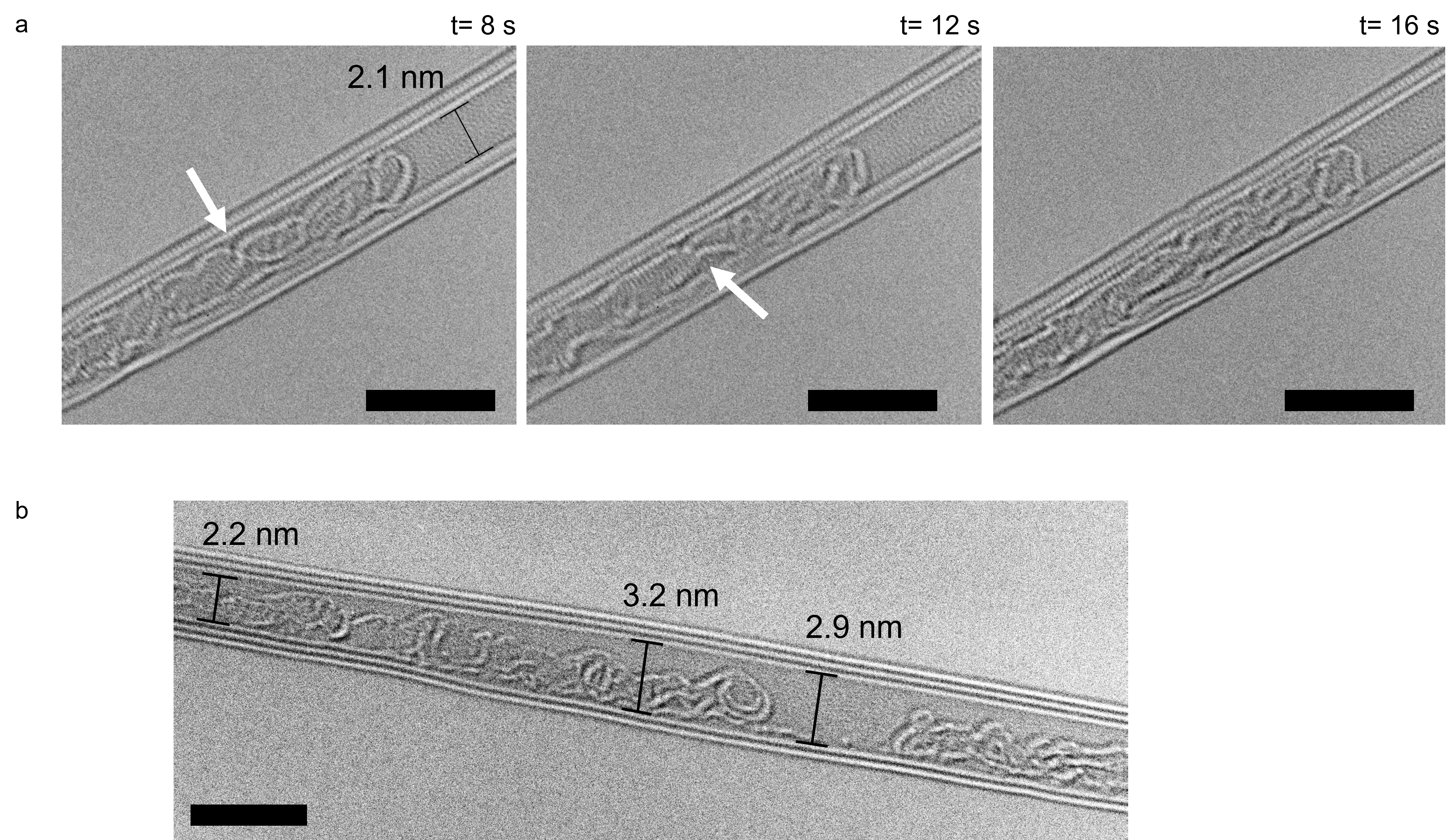}
\caption{a) GNR@BNNT twisting under the electron beam, indicated by a white arrow. b) Carbon structures formed inside a BNNT. In the small-diameter part on the left, the characteristic twist of GNRs is observed; in the wider areas in the middle, the encapsulated material forms more disordered structures; and on the right there is an empty volume (see also \textbf{Figure S2}, Supporting Information).\\ 
Low dosage HRTEM images were captured at a) 480kx and b) 800kx magnification with 336 msec exposure time for each image at parallel illumination with 80kV acceleration voltage. The length of the scale bar corresponds to 5 nm.}
\label{Figure_4}
\end{figure}

The inner diameter of the nanotubes was determined based on the TEM images. The distribution of diameters, based on observations on 31 independent BNNTs, is depicted in \textbf{Figure S3} (Supporting Information).  The mean inner diameter of the tubes is 2.35 nm with 0.96 nm standard deviation. The type of the carbon structures formed by annealing depends on the diameter of the nanotubes \cite{Chamberlain12}: an inner diameter $\sim$2 nm supports the unidirectional growth of ribbons (\textbf{Figure \ref{Figure_4}a}), while in larger tubes or in areas with larger diameter, more disordered structures are present (\textbf{Figure \ref{Figure_4}b}). A similar effect was observed earlier during the polymerization of coronene in carbon nanotubes.\cite{Barzegar16, Botka14}  Considering the van der Waals distance and the threshold for ribbon formation (\textbf{Figure S4}, Supporting Information) the tubes can accommodate ribbons with width in the range of 0.5-1.8 nm. The possible zigzag and armchair ribbons with these dimensions are presented in the Supporting Information (\textbf{Table S1}). Based on these results, the BNNTs used in the present study support AGNRs with 3$<$N$<$14 and ZGNRs with 2$<$N$<$8.

\section{Conclusion} Graphene nanoribbons were produced in the interior of BNNTs by encapsulating and annealing the liquid precursor 1,2,4-trichlorobenzene that was proven to form GNRs inside carbon nanotubes. The characteristic G-band of graphitic materials at 1598 cm$^{-1}$ appears in the Raman spectra after annealing to 500 $^{\circ}$C and 700 $^{\circ}$C. In the TCB@BNNT 700$^{\circ}$C sample, a peak at 1223 cm$^{-1}$ associated with GNRs can be observed. Wide-range optical absorption shows scattering in the NIR-UV range, while in the MIR region, vibrational modes associated with the encapsulated TCB can be seen. These signals disappear when the sample is annealed to form ribbons. In small-diameter nanotubes, ribbon-like structures were observed by HRTEM; while in higher diameter nanotubes more disordered structures were formed. The threshold between GNR and disordered carbon formation lies around 2.5 nm, indicating the widest possible GNRs to be 14-AGNR and 8-ZGNR.

\section{Experimental} 
BNNTs (SP10RP$^{11}$B Refined Powder with BN content  $>$ 99\%) were purchased from BNNT, LLC.\cite{BNNT-LLC} 1,2,4-trichlorobenzene was purchased from Sigma Aldrich. 

\textbf{Encapsulating TCB in BNNTs}
BNNTs were annealed in air for 2 h at 800 $^{\circ}$C in a quartz tube to remove BN and B$_{2}$O$_{3}$ impurities\cite{Allard20}, then connected to dynamic vacuum at 800 $^{\circ}$C for 1h to clear the interior, and let cool down at 5 $^{\circ}$C/min to room temperature. TCB was added while the system was still under vacuum. The mixture was placed in a bath sonicator (45 kHz 100 W) for 10 min, stored for 24 h and filtered. The product was left to dry in air\cite{Cadena22} for 5 days, while the TCB evaporated. 

\textbf{Preparation of GNR@BNNT} TCB@BNNT was placed in a quartz tube and connected to dynamic vacuum at room temperature for 20 minutes, to ensure the complete removal of the adsorbed molecules. Subsequently the tube  was sealed and heated, starting from a cold furnace, to 500 and 700 $^{\circ}$C, respectively, then annealed there for 12 hours. 

\textbf{Wide-range optical absorption measurements} Samples for optical (mid-infrared through ultraviolet)  measurements were prepared by pressing dry TCB@BNNT, TCB@BNNT 500 $^{\circ}$C and TCB@BNNT 700 $^{\circ}$C in KBr pellets. The spectra from the mid-infrared through the visible were measured by a Bruker Vertex V80v spectrometer. A BWtex Exemplar Plus spectrometer was used to cover the near-infrared, visible and UV range (1000-200 nm).

\textbf{Raman spectroscopy} was performed with a Renishaw InVia spectrometer using laser excitation 532, 633 and 785 nm; a Ntegra spectrometer NT-MDT equipped with a 473 nm laser; and a Renishaw InVia spectrometer equipped with a 355 nm laser.

\textbf{High resolution transmission electron microscopy}
was carried out using an aberration-corrected Thermo Fisher Scientific 200 FEI THEMIS high-resolution transmission electron microscope at 80 kV at room temperature. STEM-EDS maps were recorded using Super-X EDX detectors. The sample was sonicated in 2x ion exchanged water for 3 minute, then transferred to a lacey carbon TEM grid with glass capillary.

\medskip
\textbf{Supporting Information} \par  
Supporting Information is available from the Wiley Online Library or from the author.
 
\medskip
\textbf{Acknowledgements} \par 
Research infrastructure was provided by the Hungarian Academy of Sciences. The authors thank Zsolt Szekr\'enyes from Semilab 
Ltd. for the opportunity to use their UV Raman spectrometer. This research was funded by the Hungarian National Research, Development and Innovation Office under grant nos. FK 125063, FK 138411, TKP-2021-NVA-04 and TKP-2021-NKTA-05. A.C. was supported by the Tempus Public Foundation through a Stipendium Hungaricum Scholarship. 

\medskip

%

\begin{figure}[h]
\textbf{Table of Contents}\\
\medskip
  \includegraphics{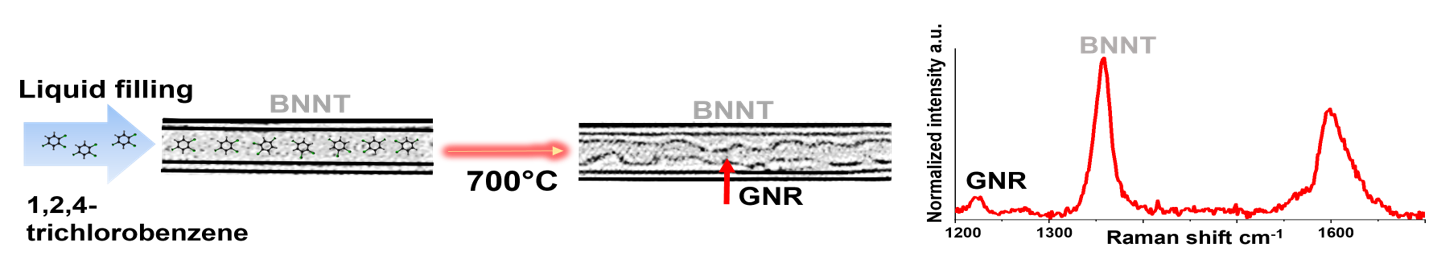}
  \medskip
  \caption*{1,2,4 trichlorobenzene was filled into boron nitride nanotubes by liquid encapsulation at room temperature. The adsorbed molecules from the outside of the nanotubes were evaporated and the encapsulated material was annealed to form graphene nanoribbons. The product was characterized by high resolution transmission electron microscopy, Raman and optical spectroscopy.}
\end{figure}

\end{document}